\documentclass[%
superscriptaddress,
reprint,
 amsmath,amssymb,
 aps,
]{revtex4-1}

\usepackage{graphicx}
\usepackage{dcolumn}
\usepackage{bm}
\usepackage[pdfencoding=auto]{hyperref}
\usepackage{bookmark}
\usepackage[mathlines]{lineno}

\usepackage[T1]{fontenc}
\usepackage{ae,aecompl}
\usepackage{xcolor} 
\usepackage{physics} 
\usepackage{tabularx}
\usepackage{booktabs}
\hypersetup{colorlinks=true,breaklinks,linkcolor=blue,urlcolor=blue,citecolor=blue}
\usepackage{placeins}

\newcommand{\deriv}[2]{\frac{\mbox{d}#1}{\mbox{d}#2}} 

\begin{document}


\title{Role of defects in ultrafast charge recombination in monolayer MoS\texorpdfstring{$_2$}{2}}

\author{Raquel \surname{Esteban-Puyuelo}}
\author{Biplab Sanyal}
 \email[]{biplab.sanyal@physics.uu.se}
 \affiliation{Division of Materials Theory, Department of Physics and Astronomy, Uppsala University, Box-516, SE 75120, Sweden}

\date{\today}

\begin{abstract}
In this work, we have systematically studied the role of point defects in the recombination time of monolayer MoS$_2$ using time-dependent \textit{ab initio} non-adiabatic molecular dynamics simulations. Various types of point defects, such as S vacancy, S interstitial, Mo vacancy and Mo interstitial have been considered. We show that defects strongly accelerate the electron-hole recombination, especially interstitial S atoms do that by 3 orders of magnitude higher compared to pristine MoS$_2$. Mo defects (both vacancy and interstitial) introduce a multitude of de-excitation pathways via various defect levels in the energy gap. The results of this study provide some fundamental understanding of  photoinduced de-excitation dynamics in presence of defects in highly technologically relevant 2D MoS$_2$. 

\end{abstract}

\maketitle


\section{Introduction}
Since the experimental realization of graphene, a single layer of graphite, in 2004 \cite{Novoselov2004}, two-dimensional materials are in the focus of solid state research. Graphene has unique characteristics and has been proposed for diverse applications, in particular, as a substitute for silicon based electronics and photovoltaics. However, the lack of a semiconducting band gap makes it unsuitable for those applications and a lot of effort has been put into finding 2D materials with similar properties and a band gap \cite{Martinazzo2010, Fujita1996, Wu2015}. One path lays in considering two-dimensional transition metal dichalcogenides (TMDs) such as MoS$_2$ \cite{Xia2014, Amani2015, Yazyev2015} because they present unique electrical and optical properties and a direct band gap. From a practical point of view, two-dimensional semiconductors have the smallest thickness possible and are thus extremely relevant in solar cells, sensing, photocatalysis and many other applications that need miniaturization. In particular, MoS$_2$ is technologically interesting because, unlike its bulk counterpart, it has a direct band gap of 1.8 eV and a high electron mobility \cite{Radisavljevic2011}.  

In applications such as photocatalysis, photo-induced excitation of electrons and hence creation of holes are important aspects of study especially the dynamics of charge-recombination. Non-radiative electron-hole recombination is one of the main channels for energy and carrier losses that reduce the material's efficiency. There are multiple experimental studies on MoS$_2$, but not many theoretical advances have been made, in particular those that consider atomic scale defects \cite{Li2017, Li2018}. Defects are very common in the fabrication of samples of TMDs, with single atom vacancies and adatoms being the most common. It is believed that defects accelerate the recombination time by introducing mid-gap energy levels. However, it should be noted that different types of atomic scale defects produce a variety of electronic structure with various sorts of defect levels in the energy gap. This is expected to have a profound influence on the dynamics of charge-recombination. To elucidate the role of these defects in charge recombination, in this project, we have studied this aspect using non adiabatic molecular dynamics (NA-MD) simulations, currently one of the few available methods that is able to provide qualitative results for periodic systems at a reasonable computational effort.

It is extremely challenging to treat the nuclear and electronic motion fully quantum mechanically for big systems containing many electrons and nuclei. Therefore, the most popular method used is a mixed quantum-classical approach where the nuclear motion is treated classically while electrons are considered in a full quantum mechanical way. Three methods have been used for non-adiabatic molecular dynamics, viz., mean field Ehrenfest dynamics, trajectory surface hopping and multiple spawning \cite{Wang2015, Crespo-Otero2018, Nelson2020, Curchod2018, Wang2016}. In the trajectory surface hopping method, a number of classical trajectories are used to approximate the evolution of the nuclear wave packet. The hopping between different Born-Oppenheimer surfaces occurs from a stochastic algorithm utilizing calculated non-adiabatic coupling parameters. Details are given in the methodology section.

In this paper, we use \textit{ab initio} NA-MD within the trajectory surface hopping approach to investigate the role of defects (S vacancy, S interstitial, Mo vacancy and Mo interstitial) in modifying the recombination time in monolayer MoS$_2$. We present the most probable charge recombination pathways for each system, by pointing out which states contribute the most to the non-adiabatic coupling and thus to the transition. The paper is organized in the following way. We introduce in \ref{section: Methods} the computational details of the electronic structure and nahNA-MD simulations, then we present our results for the ground state properties in \ref{section: GS} in absence and presence of the defects and for the non-radiative recombination in \ref{section: deexcitation}, followed by conclusions.

\section{Methodology}
\label{section: Methods}

\subsection{NA-MD calculations}
This study approaches the dynamics of electrons and nuclei in a mixed quantum-classical way: the nuclear system is treated classically via adiabatic \textit{ab initio} MD simulations, while the electronic degrees of freedom are treated quantum-mechanically. Specifically, we used NA-MD employing basis of Slater determinants composed of single-particle time-dependent (TD) adiabatic Kohn-Sham (KS) orbitals, which are computed along a given nuclear trajectory. The NA-MD trajectories were simulated using the decoherence-induced surface hopping (DISH) \cite{Tully1990, Jaeger2012} within the neglect of back reaction approximation (NBRA) \cite{Craig2005a, Fischer2011} implemented in the PYthon eXtension for Ab Initio Dynamics (PYXAID) package \cite{Akimov2013, Akimov2014}. DISH accounts for electronic decoherence by allowing  hopping between potential energy surfaces at the decoherence time. This ensures a correct partitioning of the nuclear and electronic energies. The NBRA neglects the electron-nuclear back reaction effects, relying on the fact that the nuclear dynamics is driven mainly by a ground-state potential energy surface and the electronic dynamics evolves via a parametric dependence of the non-adiabatic couplings (NAC) on the nuclear positions. This reduces the number of expensive electronic structure calculations and it has been shown to be a good approximation for large condensed matter systems where the electron-phonon dynamics is much weaker than the electron-electron couplings and the kinetic energy is the primary source of nuclear dynamics, i.e. with a rigid structure \cite{Nie2015, Nijamudheen2017, Long2016}.

The electronic properties of a system containing $N$ electrons can be obtained by solving the many-body time-dependent Schr\"odinger equation (TDSE), which for the non-relativistic case can be written as follows:
\begin{align}
i\hbar\frac{\partial\Psi}{\partial t}=\mathcal{H}\Psi, 
\label{eq: td schrodinger eq}
\end{align}
where $\mathcal{H}$ is the Hamiltonian, and $\Psi(\vec{r}_1,\vec{r}_2,...\vec{r}_N;\vec{R}_1,\vec{R}_2,...,t)$ is the all electron wavefunction, which depends on the positions of the electrons ($\vec{r}_i$) and parametrically on the ions ($\vec{R}_I$). The wavefunction can be expressed as a linear combination of adiabatic electronic wavefunctions $\psi$ and nuclear wavefunctions $\chi$ in the so-called Born-Oppenheimer expansion:
\begin{align}
\Psi(\vec{R}, \vec{r},t)=\sum_j \psi_j(\vec{r}; \vec{R} (t))\chi_j(\vec{R},t).
\end{align}
By inserting the Born-Oppenheimer expansion into \eqref{eq: td schrodinger eq}, multiplying by $\psi^*_i(\vec{r};\vec{R})$ from the left and then integrating over the electronic coordinates $\vec{r}$ we can extract the evolution of the nuclei:
\begin{align}
\left[ T_{nuc} (\vec{R)} + W_i (\vec{R}) \right]  \chi_i(\vec{R},t)+\\ 
\sum_j V_{i,j}(\vec{R})\chi_j(\vec{R},t)=i\frac{\partial}{\partial t}\chi_i ( \{\vec{R} \},t).
\label{eq: nuclei1}
\end{align}
The term $W_i(\vec{R})=E^e_i+V_{n-n}$ is the adiabatic PES for the $i$th electronic state and $V_{i,j}(\vec{R})$ is the hopping term that allows transitions between the $i$th and $j$th PES:
\begin{align}
V_{i,j}(\vec{R})=-\sum_I \frac{1}{2m_I} G_{i,j} (\vec{R})+2\vec{d}_{i,j}(\vec{R})\cdot \nabla_I,
\label{eq: 2nd order NACs}
\end{align}
where $G_{i,j} (\vec{R})=\bra{i}\nabla^2_I\ket{j}$ is the scalar coupling vector in the bra-ket notation and 
\begin{align}
\vec{d}_{i,j}(\vec{R})=\bra{i}\nabla\ket{j}
\label{eq: NACs}
\end{align}
is the derivative coupling matrix, more often called \textit{nonadiabatic coupling vector} or NAC. Since the scalar coupling vectors in eq. \eqref{eq: 2nd order NACs} are diagonal matrix elements, they can be added to the PES to redefine the energies:
\begin{align}
\varepsilon_i(\vec{R})=W_i(\vec{R})+G_{i,j} (\vec{R}).
\label{eq: PES}
\end{align}

Solving for the nuclear degrees of freedom is complicated, and one route is to use mixed quantum-classical methods. The nuclear degrees of freedom are treated classically and the electronic problem can be solved with the stationary Schr\"odinger equation, for which DFT can be used:
\begin{align}
\mathcal{H}_{el}(\vec{r},t;\vec{R})\psi(\vec{r},t;\vec{R})=W_i(\vec{R},t)\psi(\vec{r},t;\vec{R}).
\label{eq: ti schrodinger eq}
\end{align}
the electronic wavefunction $\psi(\vec{r},\vec{R},t)$ is represented in the basis of stationary adiabatic functions $\phi(\vec{r};\vec{R}(t))$ as in the following expression:
\begin{equation}
\psi(\vec{r},\vec{R},t)=\sum_i c_i(t)\phi(\vec{r};\vec{R}(t)).
\end{equation}
$c_i$ are the time-dependent expansion coefficients, and its evolution is governed by a TD Schr\"{o}dinger equation:
\begin{equation}
i\hbar\deriv{c_i}{t}=\sum_j\left(\varepsilon_i\delta_{ij}-i\hbar d_{ij}\right)c_j,
\label{eq: final pyxaid eq}
\end{equation}
where $\varepsilon$ is the diagonal part of the electron Hamiltonian defined in eq \eqref{eq: PES} and the NACs are off-diagonal terms. In practice, the NACs are calculated between two adjacent times in the  \cite{HammesSchiffer1994}, and more details about its implementation can be found elsewhere \citep{Akimov2013, Akimov2014}. 

\subsection{Fewest Switches Surface Hopping and Neglect of Back Reaction Approximation}
Eq. \eqref{eq: final pyxaid eq} describes the coherent time evolution of electronic states  coupled to the evolution of nuclear states and it can be solved with different kinds of approximations. One of them is to use surface hopping techniques, for example with the Fewest Switches Surface Hopping (FSSH) formulation by Tully \cite{Tully1990}, in which the time evolution of the nuclear wavefunctions is represented by an ensemble of trajectories that propagate via the combination of the time-dependent Schr\"odinger equation and stochastic factors. The electron population for all the states and and at all times in FSSH is equal to the one obtained from the TDSE, and the latter are given by the diagonal terms in the electronic density matrix:
\begin{align}
\rho_{ij}=c^{*}(t)c_j(t).
\end{align}
The probability for the transition from an electronic state $\ket{i}$ to a new state $\ket{j}$ in a small enough time interval $\Delta t$ can be expressed as
\begin{align}
g_{i\rightarrow j}(t)=\max\left( 0,P_{i\rightarrow j}(t) \right)
\end{align}
with 
\begin{align}
P_{i\rightarrow j}(t) \approx 2\frac{ \operatorname{Re} \left[c_i^* (t')c_j (t')\vec{d}_{ij}(t)\right]}{c_i^* (t')c_j (t')}.
\end{align}
These probabilities are compared to a uniformly distributed random number to determine if the system is to remain in the current PES or hop to the next one and nuclear velocities are rescaled along the NAC direction to maintain the total classical electron-nuclei energy. If that rescaling is not possible, the hop is rejected.

In the original FSSH, the nuclear and electronic degrees of freedom are completely coupled, so everything is updated on the fly. However, the NBRA \cite{Akimov2013} can be used to reduce the computational cost by making the approximation that the classical trajectory of the nuclei is independent of the electronic dynamics but the electronic dynamics still depends on the nuclear positions. In practice it means that the electronic problem can be solved with a series of pre-computed nuclear trajectories. Since the feedback from the electrons is not taken into account, this approximation is not valid if the electron-nuclear correlations are crucial, such as in small systems like molecules. However, it is expected to produce reasonably good results for extended solids and it is currently the most widespread method to tackle solid-state systems. In FSSH-NBRA, the hop rejection and velocity rescaling become
\begin{align}
g_{i\rightarrow j}(t) \longrightarrow g_{i\rightarrow j}(t)b_{i\rightarrow j}(t),
\end{align}
where the probability is scaled by a Boltzmann factor to account for the fact that transitions to states high up in energy are less probable:
\begin{align}
b_{i\rightarrow j}(t) \begin{cases}
											\exp \left(-\frac{E_j-E_i-E_{\hbar\omega}}{K_BT}\right)  \: & \mathrm{if} \: E_j>E_i+E_{\hbar\omega} \\
											1  \: & \mathrm{if} \: E_j\leq E_i+E_{\hbar\omega}.										
									\end{cases}
\end{align}
Here $K_B$ is the Boltzmann constant, $T$ is the temperature, and $E_{\hbar\omega}=\hbar\omega$ is the energy of the absorbed photon in case of light-matter interaction. 

\subsection{Treatment of decoherence}
By having a classical description of the nuclei, the nuclear wavefunction has been transformed into a classical phase-space point. FSSH misses the loss of quantum coherence within the electronic subsystem that is induced by the interaction with the quantum-mechanical vibrations and develop non-physical coherence. Decoherence can be introduced with \textit{decoherence-induced surface hopping} (DISH), which allows hops at the decoherence times only. These are calculated from the pure dephasing times, which are calculated in the optical response theory using the autocorrelation function of the energy gap fluctuation along the MD trajectory due to the nuclear motion, as described in Ref.  \onlinecite{Akimov2014}. 

\subsection{Computational details}
\begin{figure}[!h]
	\centering
	 \includegraphics[width=\linewidth]{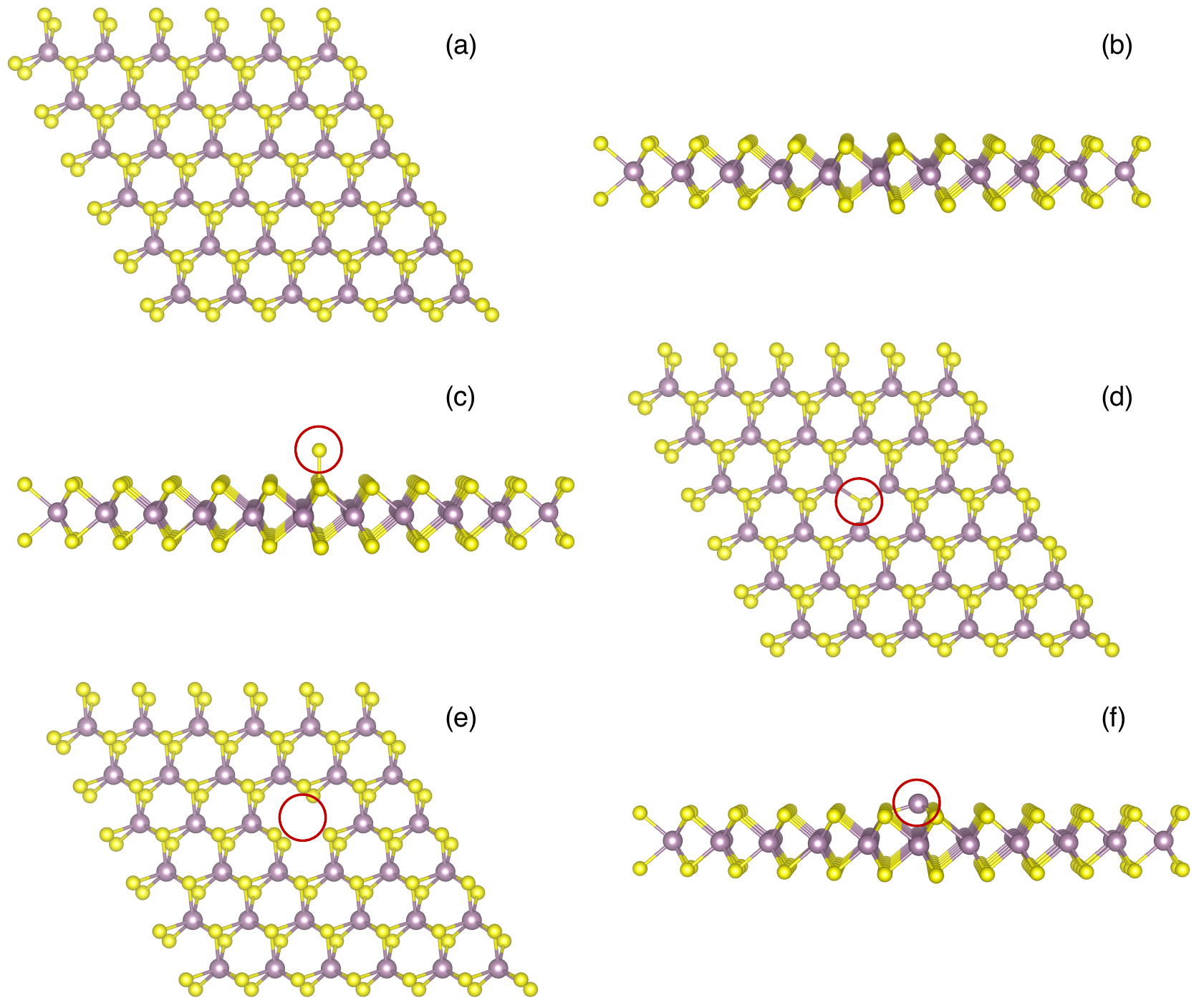}
	 \caption{\label{fig0}Supercells of the studied MoS$_2$ systems. S atoms are represented in yellow and Mo atoms in violet. (a) and (b) show top and side views of the pristine supercell respectively. (c-f) show the defected structures with the defect site marked with a red circle. (c) is int-S, (d) is vac-S, (e) is vac-Mo and (f) is int-Mo.}
	 \label{fig: structures}
\end{figure}

All the calculations have been performed with a monolayer MoS$_2$ supercell generated by repeating the primitive cell six times in the in-plane directions. A vacuum of 20 \AA\ in the out-of-plane direction is included in order to avoid spurious interaction between periodic images of the monolayer. The 6$\times$6$\times$1 supercell has been chosen in order to achieve the correct band folding: monolayer MoS$_2$ has its direct band gap in the K point of its primitive unit cell, and it can be captured by $\Gamma$ point sampling in supercells that are commensurate with the K point (multiples of 3). The pristine supercell used in this study contains 108 atoms, and we have considered point defects of Mo and S atoms, both as vacancy and interstitial adatom, as shown in figure \ref{fig: structures}. The defect concentration is thus $2.75\cdot10^{13}\ \mathrm{cm^{-2}}$, which is high compared to experimental conditions due to the need to find a compromise between modelling large supercells and its computational cost.


Geometry optimisations, adiabatic molecular dynamics simulations and electronic structure calculations were performed using the plane-wave based density functional theory based {\sc Quantum ESPRESSO} software \cite{Giannozzi2009, Giannozzi2017}. The generalized gradient approximation of Perdew, Burke and Ernzerhof (PBE) \cite{Perdew1996a, Perdew1997} was used for the exchange-correlation functional and the core electrons were represented via scalar relativistic Projector Augmented Wave pseudopotentials \cite{Kresse} with non-linear core corrections. Semicore $s$ and $p$ states were included in the valence for Mo atoms. The plane wave basis size was chosen after convergence tests and was defined by a kinetic energy cutoff of 60 Ry and a charge density cutoff of 300 Ry. The atomic coordinates of the supercells were optimized via the BFGS quasi-Newton algorithm \cite{Fletcher1987} using a fixed cell obtained from a variable cell relaxation of the MoS$_2$ unit cell. Since it has been already shown that all the systems considered are non-magnetic \citep{Haldar2015}, all calculations were done without spin polarization. 

\begin{figure}[h!]
	\centering
 	\includegraphics[width=\linewidth]{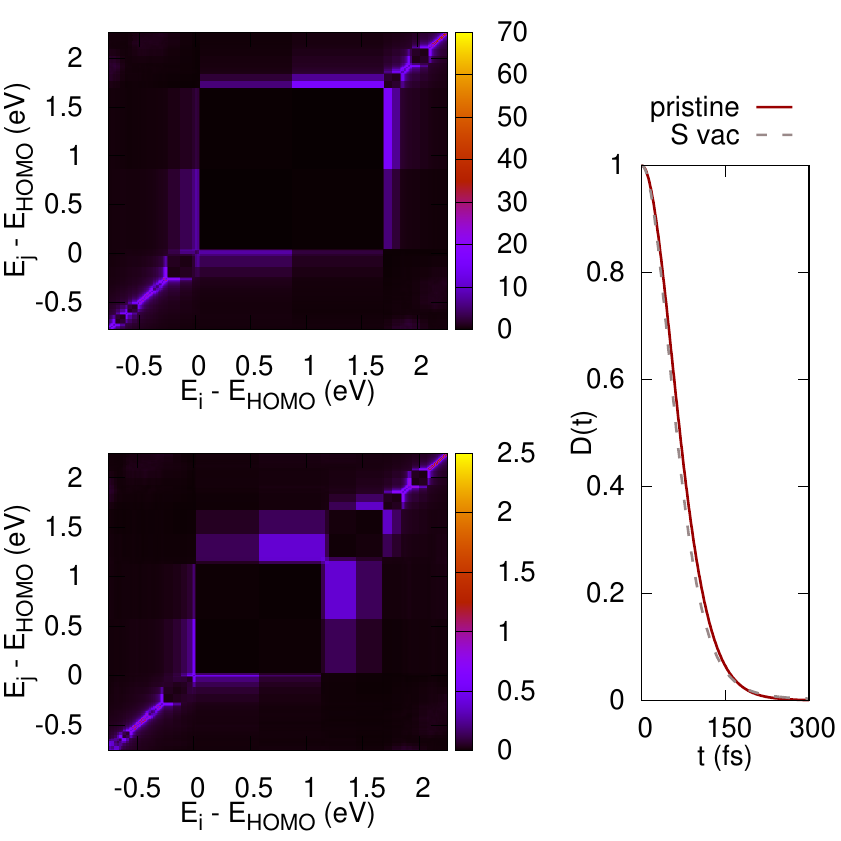}
 	\caption{\label{fig6}Left: NAC matrices for pristine (top) and vac-S (bottom) MoS$_2$. The 0 in the horizontal axis is the VBM and positive values stand for levels above the VBM (unoccupied) whereas negative values are for levels below the VBM (occupied). The vertical axis is defined in the same way as for the CBM. The bright color indicates the magnitude of the NAC in meV. Right: decoherence function for the CBM-VBM transition for both systems. The decoherence time is extracted from an exponential fitting of this function.}
\end{figure}

Nuclear trajectories were produced by running a 2.5 ps ground state adiabatic MD simulations using the Verlet algorithm \cite{Verlet1967} with 1 fs time step and the Andersen thermostat \citep{Andersen1980} was used to maintain ambient temperature (300 K). At each time step, the electronic problem was solved self-consistently at the $\Gamma$ point. The NACs were calculated for a subset of 60 KS orbitals (30 occupied and 30 unoccupied) falling inside a 4 eV window, reasonable for optical excitations. An example of how these matrices look like is shown in figure \ref{fig6} for pristine and vac-S MoS$_2$.

The NA-MD trajectories were calculated for 50 initial conditions sampled from the precomputed adiabatic MD trajectory, and for each initial condition, 10000 stochastic realizations of the surface hopping process were computed. By averaging all initial conditions and stochastic realizations, the evolution of the populations of all computed states was obtained. The relaxation time was then extracted by fitting an exponential function to the population.

\section{Results and discussions}
\subsection{Ground state properties}
\label{section: GS}
In this section we discuss the electronic properties of pristine and defected MoS$_2$ by analyzing the total density of states (DOS) and charge densities.

\begin{figure*}
 \includegraphics[width=\linewidth]{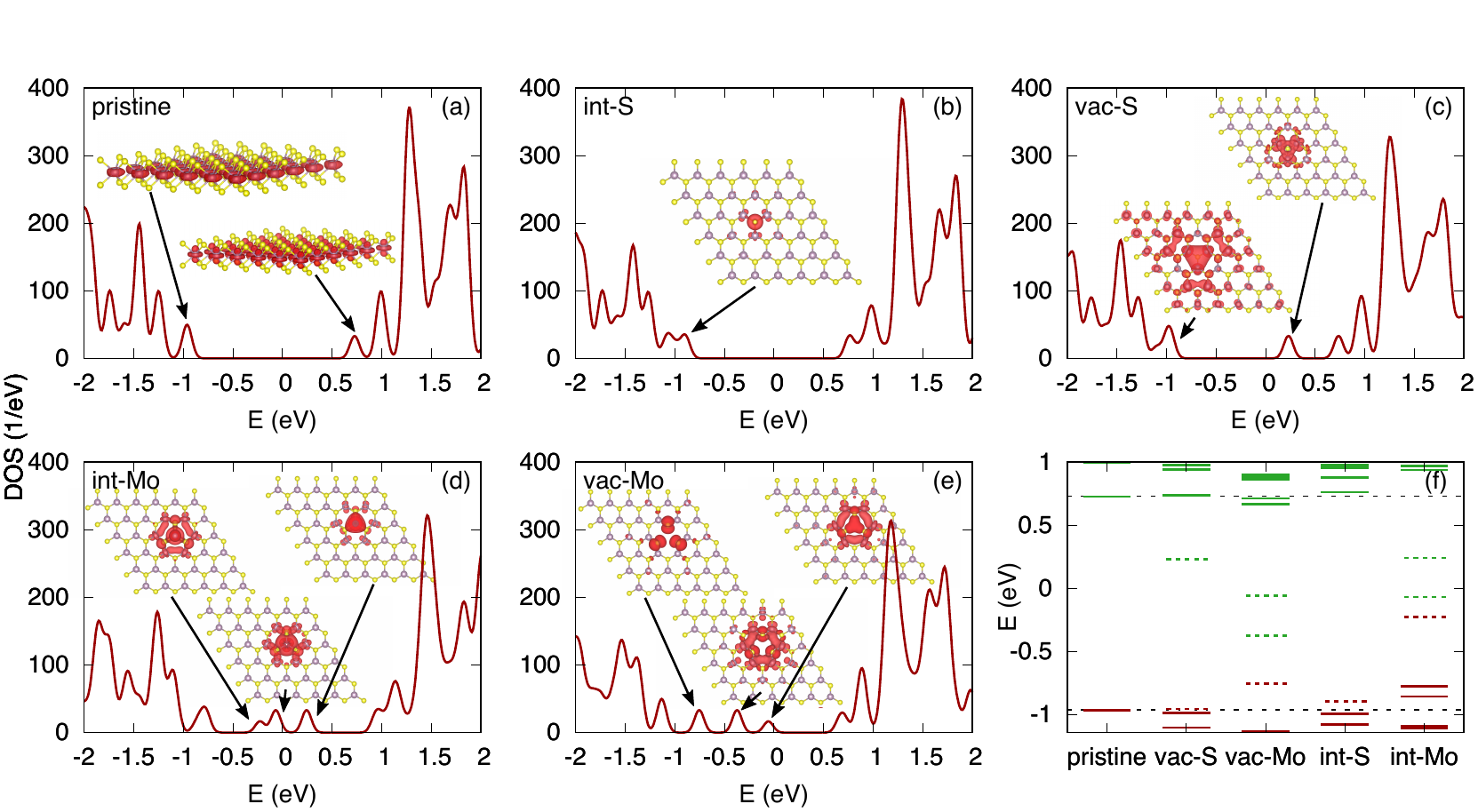}
 \caption{\label{fig1} (a)-(e) Density of states for the supercells considered. (a) Charge density of the pristine CBM and VBM as insets. (b)-(e) Charge density of the defect states that appear inside the pristine gap in each of the defected systems. (f) Energy level diagram for the pristine and different defected structures in MoS$_2$. The red and green horizontal lines represent occupied and unoccupied energy levels respectively, and the dashed red and green lines indicate that those energy levels belong to occupied and unoccupied defect states respectively inside the gap. The dashed black lines mark where the CBM and the VBM are located in the pristine system.}
 \label{fig: dos}
\end{figure*}

Figure \ref{fig: dos} (f) shows an energy level diagram for the systems studied in this project: pristine, S vacancy (vac-S), S interstitial (int-S), Mo vacancy (vac-Mo) and Mo interstitial (int-Mo). The dashed lines show the position of the valence band maximum (VBM) and conduction band minimum (CBM) of the pristine system, while the short colored lines denote the actual positions of the energy levels. The occupied levels are shown in red and the unoccupied ones in green. The DOS of pristine MoS$_2$ is shown in figure \ref{fig: dos} (a), where the charge density of the CBM and VBM are also shown. They match with the calculated charge density at the k-point K of a unit cell, as well as to previous studies \cite{Autieri2017} so the states are well captured in the supercell. 

By creating defects, new states are created, mainly in the band gap, and their charge densities are plotted in figure \ref{fig: dos} panels (b) to (e) for each defected structure. Further analysis of the DOS and charge densities shows that the character of the CBM and VBM is maintained to a relatively good extent in the defected structures and therefore their charge densities are not plotted in figure \ref{fig: dos}. By including an interstitial S adatom (int-S), an occupied defect level appears lying very close to the CBM but is very localized, as seen in the charge density in figure \ref{fig: dos} (b). It is worth noticing that the presence of this defect produces an appreciable distortion in the character of the system's CBM, which will have an effect in the dynamics, as it will become apparent later. The next defect we considered is a S atom vacancy (vac-S), which creates two defect levels: an occupied defect level extremely close to the CBM and a deep unoccupied defect level. Their DOS and charge densities are included in figure \ref{fig: dos} (c). Mo defects create more complex energy diagrams. Since Mo atoms have more electrons than S, the energy shift of the VBM and CBM, upwards in the case of an interstitial adatom and downwards in the vacancy case, which are larger than in the case of S. Furthermore, Mo vacancy (vac-Mo) creates an occupied state close to the CBM, which induces a distortion in the CBM, and two unoccupied levels in the middle of the gap (figure \ref{fig: dos} (e)). An interstitial Mo adatom (int-Mo) creates an occupied and two unoccupied defect levels, all three in the middle of the gap (\ref{fig: dos} (f)). These results are in agreement with previous studies \cite{Haldar2015}.

\subsection{De-excitation dynamics}
\label{section: deexcitation}
\begin{figure}[!h]
 \includegraphics[width=\linewidth]{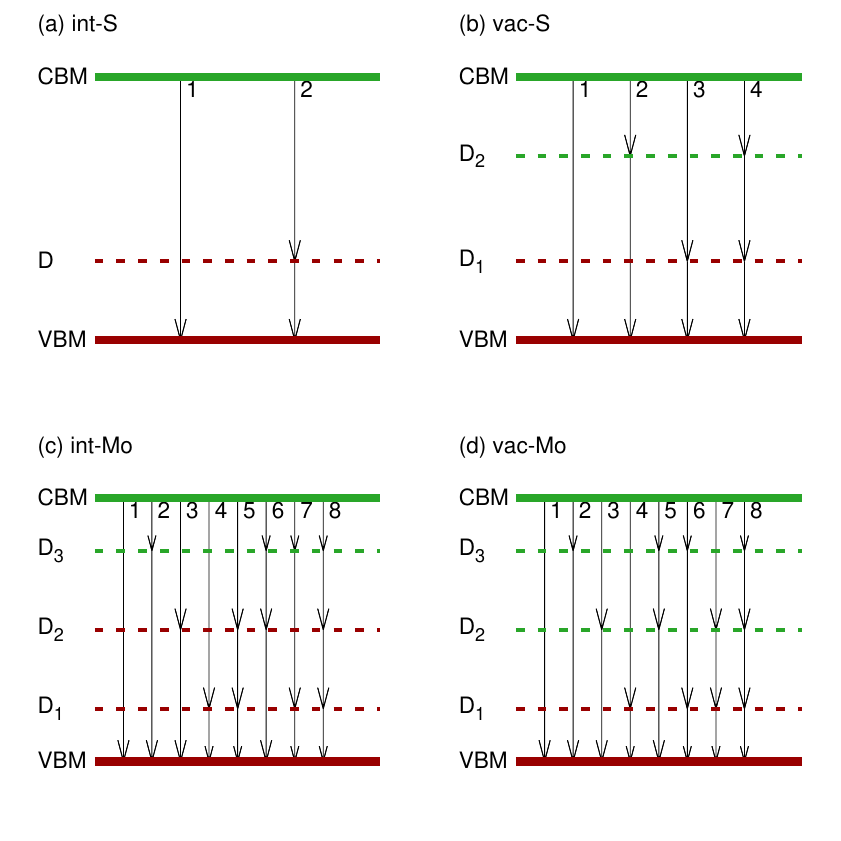}
 \caption{Scheme of the possible transitions in (a) int-S, (b) vac-S, (c) int-Mo and (d) vac-Mo. Red and green lines indicate occupied and unoccupied levels, respectively. Defect states inside the gap are shown with a dashed line and, within each system, they are denoted as D$_\mathrm{n}$ with $\mathrm{n}=1,2,3$ with increasing energy.}
 \label{fig: transitions}
\end{figure}

For each system, we selected the energy levels that can participate in the recombination in order to form the \textit{active space} in which the DISH algorithm will be applied. For instance, in the pristine system, the VBM and CBM are doubly degenerate so the active space is formed by 4 orbitals and we consider excited states corresponding to the CBM $\rightarrow$ VBM transition, where the population of the excited state is expected to decay to the ground state. To obtain the recombination timescale $\tau$, we fit the population increase in the ground state with an exponential function of the form
\begin{align}
P(t)=1-\exp\left( \frac{-t}{\tau}  \right),
\end{align}
so that it reaches the normalized value of 1 at infinite time. For the defected systems, many intermediate transitions need to be considered, and a scheme of the energy levels is shown in figure \ref{fig: transitions}. The unoccupied defects act as electron traps because excited electrons coming from the CBM can fall in these defect levels before recombining with the holes in the VBM. In the opposite way, occupied defect states act as hole traps. We consider all the transitions with all the intermediate steps. The calculated transition timescales are presented in tables \ref{tab: tau pristine}-\ref{tab: tau vac-Mo}. For each system, we provide the timescales for the direct recombination and each alternative mechanism that includes transitions involving defect states. For the latter, the timescales of all the subprocesses are shown separated by commas. This is due to the fact that, although there is a clear expression for the effective time for parallel transitions (addition of the transition rates), there is no general analytical expression for series transitions. This is the case of each of the alternative transitions to direct recombination, as for instance, in int-S, the process denoted by the term 'Hole trap' in table \ref{tab: tau int-S} is in fact a series of transitions CBM$\rightarrow$D$\rightarrow$VBM (see the corresponding figure \ref{fig: transitions} a).

\begin{table}[!h]
\caption{\label{tab: tau pristine}Recombination times and mechanism in pristine MoS$_2$.}
\begin{center}
\begin{tabular}{lcc} \toprule
Process & Mechanism & $\tau$ (ps) \\
\hline
1: Direct recombination & CBM $\rightarrow$ VBM & 72876 \\
\bottomrule
\end{tabular}
\end{center}
\end{table}

\begin{table}[!h]
\caption{\label{tab: tau int-S}Recombination times and mechanisms in int-S MoS$_2$. The fastest mechanism appears in bold. For the mechanisms involving defect states, the timescales of the subprocesses are separated by comas.}
\begin{center}
\begin{tabular}{lcc} \toprule
Process & Mechanism & $\tau$ (ps) \\
\hline
1: Direct recombination & CBM $\rightarrow$ VBM & 21139 \\
\textbf{2: Hole trap} & CBM $\rightarrow$ D $\rightarrow$ VBM & 6, 44 \\
\bottomrule
\end{tabular}
\end{center}
\end{table}

\begin{table*}
\caption{\label{tab: tau vac-S}Recombination times and mechanisms in vac-S MoS$_2$. The fastest mechanism is shown in bold. For the mechanisms involving defect states, the timescales of the subprocesses are separated by comas.}
\begin{center}
\begin{tabular}{lcc} \toprule
Process & Mechanism & $\tau$ (ps) \\
\hline
1: Direct recombination & CBM $\rightarrow$ VBM & 67784 \\
\textbf{2: Electron trap} & CBM $\rightarrow$ D$_2$ $\rightarrow$VBM & 406, 3440 \\
3: Hole trap & CBM $\rightarrow$ D$_1$ $\rightarrow$ VBM & 71700, 1 \\
4: Electron and hole traps & CBM $\rightarrow$ D$_2$ $\rightarrow$ D$_1$ $\rightarrow$ VBM & 406, 2978, 1 \\
\bottomrule
\end{tabular}
\end{center}
\end{table*}

\begin{table*}
\caption{\label{tab: tau int-Mo}Recombination times and mechanisms in int-Mo MoS$_2$. The fast pathways are shown in bold. For the mechanisms involving defect states, the timescales of the subprocesses are separated by comas.}
\begin{center}
\begin{tabular}{lcc} \toprule
Process & Mechanism & $\tau$ (ps) \\
\hline
1: Direct recombination & CBM $\rightarrow$ VBM & 52418 \\
2: Electron trap & CBM $\rightarrow$ D$_3$ $\rightarrow$ VBM & 180, 3251 \\
\textbf{3: Hole trap 1} & CBM $\rightarrow$ D$_2$ $\rightarrow$ VBM &  217, 16\\
4: Hole trap 2 & CBM $\rightarrow$ D$_1$ $\rightarrow$ VBM & 4429, 290\\
\textbf{5: Double hole trap} & CBM $\rightarrow$ D$_2$ $\rightarrow$ D$_1$ $\rightarrow$ VBM & 217, 2, 290 \\
\textbf{6: Electron and hole traps 1} & CBM $\rightarrow$ D$_3$ $\rightarrow$ D$_2$ $\rightarrow$ VBM & 180, 26, 16 \\
7: Electron and hole traps 2 & CBM $\rightarrow$ D$_3$ $\rightarrow$ D$_1$ $\rightarrow$ VBM & 180, 4429, 290 \\
8: Electron and double hole traps & CBM $\rightarrow$ D$_3$ $\rightarrow$ D$_2$ $\rightarrow$ D$_1$ $\rightarrow$ VBM & 180, 26, 2, 290 \\
\bottomrule
\end{tabular}
\end{center}
\end{table*}

\begin{table*}
\caption{\label{tab: tau vac-Mo} Recombination times and mechanisms in vac-Mo MoS$_2$. The fast mechanisms are shown in bold. For the mechanisms involving defect states, the timescales of the subprocesses are separated by comas.}
\begin{center}
\begin{tabular}{ccc} \toprule
Process & Mechanism & $\tau$ (ps) \\
\hline
1: Direct recombination & CBM $\rightarrow$VBM & 32284 \\
2: Electron trap 1 & CBM $\rightarrow$D$_3$ $\rightarrow$ VBM & 161, 742 \\
3: Electron trap 2 & CBM $\rightarrow$ D$_2$ $\rightarrow$ VBM & 1436, 74 \\
\textbf{4: Hole trap} & CBM $\rightarrow$ D$_1$ $\rightarrow$ VBM & 370, 116 \\
5: Double electron trap & CBM $\rightarrow$ D$_3$ $\rightarrow$ D$_2$ $\rightarrow$ VBM & 161, 1436, 116 \\
\textbf{6: Hole and electron traps 1} & CBM $\rightarrow$ D$_3$ $\rightarrow$ D$_1$ $\rightarrow$ VBM & 161, 46, 116 \\
7: Hole and electron traps 2 & CBM $\rightarrow$ D$_2$ $\rightarrow$ D$_1$ $\rightarrow$ VBM & 1436, 46, 116 \\
8: Hole and double electron traps & CBM $\rightarrow$ D$_3$ $\rightarrow$ D$_2$ $\rightarrow$ D$_1$ $\rightarrow$ VBM & 161, 32, 370, 116\\
\bottomrule
\end{tabular}
\end{center}
\end{table*}


The recombination time for pristine MoS$_2$ is, according to our calculations, around 70 ns, as table \ref{tab: tau pristine} shows. Since non-radiative electron-hole recombination is the main source of energy losses in electronic devices and solar cells, this very long non-radiative recombination time is promising, since it means that efficient devices can be built from very pure MoS$_2$ samples. However, defects clearly accelerate recombination. For instance, in systems with S adatoms (int-S, see table \ref{tab: tau int-S} and figure \ref{fig: transitions} a) the direct recombination time is reduced by a factor 3. This is due to the presence of the localized defect level lying close to the CBM, which distorts the CBM orbital character compared to the pristine case. Additionally, the defect level acts as a hole trap and seems to be a much faster pathway, since the partial transitions (CBM $\rightarrow$ D and D $\rightarrow$ VBM) happen in the ps timescale. It has been shown theoretically in \cite{Haldar2015} that these defects have the lowest formation energy of all the discussed ones (around 1 eV), and they are also the most common defect in experimental samples grown by physical vapor deposition \cite{Hong2015}. S vacancies (vac-S), which are present in chemical vapor deposition samples \cite{Zhou2013}, are not so detrimental to the recombination time, as shown in table \ref{tab: tau vac-S}. In fact, the direct recombination behaves in a very similar way as in the pristine sample, due to the fact that the defect levels do not distort the original CBM and VBM. It is worth noticing that the transition D$_1$ $\rightarrow$ VBM (see figure \ref{fig: transitions} b) is remarkably fast (1ps), which means that the CBM and this closely lying delocalized level can be considered to overlap. On the other hand, the unoccupied defect level D$_2$ acts as an electron trap that seems to be the fastest transition mechanism, with a timescale that is about half of the direct path. 

Mo defects have higher formation energy and they are therefore less frequently observed in experiments, but it is anyhow interesting to study their relaxation dynamics. The direct recombination in systems with interstitial Mo defects (int-Mo, see table \ref{tab: tau int-Mo}) is reduced only by a factor $1/3$ compared to the pristine case, which can be explained by the fact that its CBM and VBM are not appreciably distorted by the presence of the multiple defect levels. Additionally, it seems that the occupied defect D$_2$ acts as a hole trap (mechanism 3) that accelerates the recombination to timescales in the order of few hundreds of ps. Mechanisms 6 and 7 include multiple traps and are competing pathways and the relaxation dynamics in this system will be a combination of these three mechanisms. Mo vacancies (vac-Mo, table \ref{tab: tau vac-Mo}) accelerate the direct recombination by half, which is coherent with the slight distortion of the CBM induced by the defects, as it happens in the int-S system. However, the occupied defect level D$_1$ acts as a hole trap (mechanism 1) and has a transition timescale of a few hundreds of ps, which competes with mechanism 6 (hole and electron traps 1), which includes the unoccupied D$_3$ and has a similar timescale.

It is clear that the presence of point defects accelerates the direct recombination in monolayer MoS$_2$, in particular int-S does so by reducing the time to 1/3 of the pristine one. However, the same defect states create alternative pathways that are always faster than their respective direct recombinations. Table \ref{tab: MoS2 results} contains a summary of the most probable mechanisms for each studied system. 
\begin{table*}
\caption{\label{tab: MoS2 results} Summary of the fastest pathways in each of the studied systems. The numbers in parentheses are related to the transitions in Figure \ref{fig: transitions} and the respective tables \ref{tab: tau pristine}-\ref{tab: tau vac-Mo}}
\begin{center}
\begin{tabular}{cc} \toprule
System  & Processes \\
\hline
pristine  & Direct (1) \\
int-S  & Hole trap (2) \\
vac-S  & Electron trap (2) \\
int-Mo  & Hole trap 1 (3), double hole trap (5), and electron and hole traps 1 (6) \\
vac-Mo  & Hole trap (4), and  hole and electron traps 1 (6)\\
\bottomrule
\end{tabular}
\end{center}
\end{table*}

In order to compare our results with experiments we can make use of the quantum yield, which is a measure of how photo-efficient a system is. In our case it can be related to the radiative $\tau_R^-{1}$ and non-radiative $\tau_{NR}^{-1}$ recombination rates:
\begin{align}
QY=\frac{\tau_R^{-1}}{\tau_R^{-1}+\sum \tau_{NR}^{-1}}.
\label{eq: QY}
\end{align}
The radiative processes involve photon emission and are typical for direct semiconductors. We distinguish between two non-radiative processes: Shockley-Read-Hall and Auger effects. In the first one, the recombination is assisted by a trap (mid-gap level) and can happen at low carrier concentrations both in direct and indirect semiconductors, in which lattice vibrations play an important role. Auger recombination is predominant at high carrier concentrations. The timescales that we compute with our method are non-radiative recombination times $\tau_{NR}$, and we can use \eqref{eq: QY} to compare with photoluminiscence data, which are available mostly for pristine samples. For instance, Ref. \onlinecite{Amani2015} reports a $QY$ of nearly 95\% which gives a non-radiative recombination time of 11 ns at room temperature, in qualitative agreement with our calculations of 70 ns in the pristine case, and even closer in the presence of defects that reduce the effective recombination time. We have obtained longer recombination times than the ones reported in other theoretical works \cite{Li2017, Li2018}, although we agree in the fact that S adatom defects (int-S) result in much faster recombination times than S vacancies (vac-S). Plausible reasons for our longer timescales could be their treatment of the long term dynamics and higher density of defects due to their smaller supercell. Another explanation to the fact that we obtain longer times than the observed in experiments are excitons. It is known that excitonic effects can accelerate the recombination dynamics \cite{Smith2021, Smith2021a}, and MoS$_2$ has the well known A and B excitons. However, we would need to go beyond the one-particle picture we have used in this study to be able to capture those in a quantitative way, and it would be computationally prohibitive. Nevertheless, the one-particle picture is a fair approximation for the cases we have studied, as we have considered only the lowest excited states and they do not mix a large number of Slater determinants, see the discussions in \onlinecite{Smith2021} and \onlinecite{Izmaylov2008}. Therefore, we are able to describe non-radiative recombination dynamics in monolayer MoS$_2$ with defects in a qualitative way.

We have observed that the exact transition times are sensitive to the decoherence scheme that we use, which is consistent with other similar studies (see this topics discussed in the recent review \onlinecite{Smith2020} ad references therein). 
This requires special attention and will be studied in detail in near future. Furthermore, the computational method that we use in PYXAID does not enforce that the wavefunctions have the same phase at consecutive time steps, which would be necessary in the NAC calculation \cite{Akimov2018a}. The phase inconsistency has been addressed in a recent development of Libra library \cite{Akimov2016}, which will be explored in the near future.




\section{Conclusion}
In this paper, we have shown that point defects (S and Mo adatoms and vacancies) accelerate the effective recombination time in monolayer MoS$_2$ samples by using NA-MD simulations. 
Defects have two effects in the recombination times: first, they reduce the direct CBM$\rightarrow$VBM time by as much as half compared to the pristine case; second and more importantly, they introduce mid-gap defect levels that allow for faster pathways for the recombination, in some case in the order of few hundreds of ps. Our calculated non-radiative recombination time of around 70 ns for pristine MoS$_2$ qualitatively agrees with what has been reported experimentally. Defects, specially those involving S atoms, are rather common in experimental samples of monolayer MoS$_2$. However, even if they have higher formation energy, Mo defects can be introduced controllably using helium ion beam \cite{Aryeetey2020}. We have shown that Mo defects introduce many more alternative pathways that accelerate recombination through the defect states, compared to S vacancies and adatoms, which has not been reported before. In particular, we predict that both Mo adatoms (int-Mo) and vacancies (vac-S) can accelerate the recombination in the order of hundreds of ps, and therefore are more detrimental to the material efficiency than S vacancies. In summary, we have analyzed the mechanisms and timescales through which non-radiative recombination can happen in realistic samples of monolayer MoS$_2$, which is extremely important in order to move forward in the development of technology based on this material.


\section*{Acknowledgments}
We are thankful to Alexey V. Akimov from SUNY Buffalo (USA) for introducing us to PYXAID and helping at various steps regarding technicalities and analysis. We thank Oscar Gr\aa n\"as for useful discussions. BS acknowledges financial support from Swedish Research Council (grant 2016-05366 and grant 2017-05447) and STINT initiation grant program (IB2015-6063). REP thanks the financial support from the strategic research council (SSF) grant ICA16-0037. We are grateful to NSC under Swedish National Infrastructure for Computing (SNIC), the PRACE-2IP project resource Salomon cluster based in Czech Republic at the IT4Innovations and DECI-15 project DYNAMAT for allocation of supercomputing time.

\end{document}